\pdfoutput=1

\documentclass[twocolumn,prd,reprint,preprintnumbers,amsmath,amssymb,superscriptaddress,nofootinbib]{revtex4}

\usepackage{aas_macros}
\usepackage{epsfig,psfrag,graphics,verbatim}
\usepackage{graphicx}
\usepackage{dcolumn}
\usepackage{bm}
\usepackage{color}
\usepackage{multirow}
\usepackage{url}
\usepackage{hyperref}
\usepackage{enumitem}
\usepackage[normalem]{ulem}

\newcounter{bla}

\begin{document}

\title{\texttt{galkin}: a new compilation of the Milky Way rotation curve data}

\author{Miguel Pato}
\affiliation{The Oskar Klein Centre for Cosmoparticle Physics, Department of Physics, Stockholm University, AlbaNova, SE-106 91 Stockholm, Sweden}

\author{Fabio Iocco}
\affiliation{ICTP South American Institute for Fundamental Research, and Instituto de F\'isica Te\'orica - Universidade Estadual Paulista (UNESP), Rua Dr.~Bento Teobaldo Ferraz 271, 01140-070 S\~{a}o Paulo, SP Brazil}

\date{\today}

\begin{abstract}
We present \texttt{galkin}, a novel compilation of kinematic measurements tracing the rotation curve of our Galaxy, together with a tool to treat the data. The compilation is optimised to Galactocentric radii between $3$ and $20\,$kpc and includes the kinematics of gas, stars and masers in a total of 2780 measurements carefully collected from almost four decades of literature. A simple, user-friendly tool is provided to select, treat and retrieve the data of all source references considered. This tool is especially designed to facilitate the use of kinematic data in dynamical studies of the Milky Way with various applications ranging from dark matter constraints to tests of modified gravity.
\end{abstract}

\maketitle

\section{Introduction}
\par The rotation curve of a spiral galaxy provides far-reaching insight into its properties, as noticeably explored for decades now (see e.g.~Refs.~\cite{1970ApJ...160..811F,1973A&A....26..483R,1978PhDT.......195B,1978ApJ...225L.107R, Persic:1995tc}). Data on the rotation curve of our Galaxy, a spiral itself, have also been available for several decades \cite{Blitz1979,BurtonGordon1978,Clemens1985,Knapp1985,Fich1989,Pont1994}. However, the data are rather disperse throughout the literature and groups of references are often neglected. We therefore set out to assemble a comprehensive compilation of the decades-long observational effort to pinpoint the rotation curve of the Milky Way. The compilation, named \texttt{galkin}\footnote{To download your copy of \texttt{galkin},  please refer to our GitHub page  \href{https://github.com/galkintool/galkin}{\tt github.com/galkintool/galkin} or contact us at \href{mailto:galkin.tool.mw@gmail.com}{galkin.tool.mw@gmail.com}.}, improves upon existing ones (e.g.~Refs.~\cite{Sofue2009,Bhattacharjee2014}) on several aspects, including most notably: (i) an enlarged database of observations appropriately treated for unified use, and (ii) the release of a simple out-of-the-box tool to retrieve the data. This compilation has been presented in Ref.~\cite{2015NatPh..11..245I} and later adopted in other works in the literature. Without venturing into any analysis of the Galactic structure or dynamics (as done in \texttt{galpy} \cite{2015ApJS..216...29B}), here we provide instead a thorough description of the data sets (Sec.~\ref{sec:comp}) as well as the features of an out-of-the-box tool (Sec.~\ref{sec:tool}) to access the database and output the desired data for independent analyses. The open source code provided is simple, flexible and can be easily modified to include new datasets or other types of measurements. The latter feature is particularly relevant on the eve of the precision era soon to be introduced by the Gaia satellite \cite{2012Ap&SS.341...31D} and an array of optical and near-infrared ground-based surveys such as APOGEE-2 \cite{apogee2site,APOGEE}, GALAH \cite{2015MNRAS.449.2604D}, WEAVE \cite{weavesite} and 4MOST \cite{2012SPIE.8446E..0TD}. Our compilation can be regarded as a step forward in unifying the current state of the art, yet it is certainly susceptible of further inclusions -- please see our own extensive caveats and notes throughout the manuscript. 
We encourage the community to adopt \texttt{galkin} and participate in its extension as new datasets arise.

\begin{table*}[htp]
\begin{center}
\begin{tabular}{|c|l c  c c|} 
\cline{2-5}
\multicolumn{1}{l|}{}			&	tracer type 		       							& $R$ [kpc]					& quadrants 	& $\quad$tracers$\quad$	\\
\hline
\multirow{19}{*}{gas kinematics} 	& HI terminal velocities 								&						& 		&	 					\\
					& $\quad$ Fich+ '89 \cite{Fich1989} 							& 2.1 -- 8.0					& 1,4		& 149/149	 				\\
					& $\quad$ Malhotra '95 \cite{Malhotra1995} 						& 2.1 -- 7.5				 	& 1,4		& 110/110  							\\
					& $\quad$ McClure-Griffiths \& Dickey '07 \cite{McClure-GriffithsDickey2007} 		& 2.8 -- 7.6				 	& 4		& 701/761 							\\
					& HI thickness method									&						&		&				\\
					& $\quad$ Honma \& Sofue '97 \cite{HonmaSofue1997} 					& \hspace{0.16cm}6.8 -- 20.2 			& --		& \hspace{0.16cm}13/13\hspace{0.16cm} 			\\
					& CO terminal velocities								&						& 		&					\\
					& $\quad$ Burton \& Gordon '78 \cite{BurtonGordon1978}  				& 1.4 -- 7.9				 	& 1		& 284/284  							\\
					& $\quad$ Clemens '85 \cite{Clemens1985}  						& 1.9 -- 8.0					& 1		& 143/143  						\\
					& $\quad$ Knapp+ '85 \cite{Knapp1985}  							& 0.6 -- 7.8				 	& 1		& \hspace{0.16cm}37/37\hspace{0.16cm} 					\\
					& $\quad$ Luna+ '06 \cite{Luna2006} 							& 2.0 -- 8.0				 	& 4		& 272/457  						\\
					& HII regions										&						&		&					\\
					& $\quad$ Blitz '79 \cite{Blitz1979} 							& \hspace{0.16cm}8.7 -- 11.0 			& 2,3		& \hspace{0.32cm}3/3\hspace{0.32cm} 					\\
					& $\quad$ Fich+ '89 \cite{Fich1989}  							& \hspace{0.16cm}9.4 -- 12.5 			& 3		& \hspace{0.32cm}5/104  					\\
					& $\quad$ Turbide \& Moffat '93 \cite{TurbideMoffat1993}   				& 11.8 -- 14.7 					& 3		& \hspace{0.32cm}5/8\hspace{0.32cm} 			\\
					& $\quad$ Brand \& Blitz '93 \cite{BrandBlitz1993} 					& \hspace{0.16cm}5.2 -- 16.5 			& 1,2,3,4	& 148/206			 			\\
					& $\quad$ Hou+ '09 \cite{Hou2009}  							& \hspace{0.16cm}3.5 -- 15.5 			& 1,2,3,4	& 274/815						\\
					& giant molecular clouds								&						&		&					\\
					& $\quad$ Hou+ '09 \cite{Hou2009}							& \hspace{0.16cm}6.0 -- 13.7 			& 1,2,3,4	& \hspace{0.16cm}30/963  					\\
\hline
\multirow{10}{*}{star kinematics}	& open clusters $\dagger$										&						&		&					\\
					& $\quad$ Frinchaboy \& Majewski '08 \cite{FrinchaboyMajewski2008}  			& \hspace{0.16cm}4.6 -- 10.7 			& 1,2,3,4	& \hspace{0.16cm}60/71\hspace{0.16cm} 	\\
					& planetary nebulae									&						& 		&					\\
					& $\quad$ Durand+ '98 \cite{Durand1998} 						& \hspace{0.16cm}3.6 -- 12.6 			& 1,2,3,4	& \hspace{0.16cm}79/867 			\\
					& classical cepheids									&						&		&					\\
					& $\quad$ Pont+ '94 \cite{Pont1994} 							& \hspace{0.16cm}5.1 -- 14.4 			& 1,2,3,4	& 245/278						\\
					& $\quad$ Pont+ '97 \cite{Pont1997} 							& 10.2 -- 18.5 					& 2,3,4		& \hspace{0.16cm}32/48\hspace{0.16cm} 					\\
					& carbon stars										&						&		&					\\
					& $\quad$ Demers \& Battinelli '07 \cite{DemersBattinelli2007} 				& \hspace{0.16cm}9.3 -- 22.2 			& 1,2,3		& \hspace{0.16cm}55/103 				\\
					& $\quad$ Battinelli+ '13 \cite{Battinelli2013} 					& 12.1 -- 24.8 					& 1,2		& \hspace{0.16cm}35/36\hspace{0.16cm} 				\\
\hline
\multirow{6}{*}{masers}	& masers $\dagger$											&						&		&					\\
					& $\quad$ Reid+ '14 \cite{Reid2014}  							& \hspace{0.16cm}4.0 -- 15.6 			& 1,2,3,4	& \hspace{0.16cm}80/103					\\
					& $\quad$ Honma+ '12 \cite{Honma2012} 							& 7.7 -- 9.9				 	& 1,2,3,4	& \hspace{0.16cm}11/52\hspace{0.16cm} 					\\
					& $\quad$ Stepanishchev \& Bobylev '11 \cite{StepanishchevBobylev2011} 			& 8.3						& 3		& \hspace{0.32cm}1/1\hspace{0.32cm} 					\\
					& $\quad$ Xu+ '13 \cite{Xu2013} 							& 7.9 						& 4		& \hspace{0.32cm}1/30\hspace{0.16cm} 					\\
					& $\quad$ Bobylev \& Bajkova '13 \cite{BobylevBajkova2013} 				& 4.7 -- 9.4				 	& 1,2,4		& \hspace{0.32cm}7/31\hspace{0.16cm} 					\\
\hline
\end{tabular}

\caption{The list of all kinematic measurements of the Milky Way included in \texttt{galkin}. For each reference, the range of Galactocentric radius is reported assuming $R_0=8\,$kpc along with the Galactic quadrant(s) covered and the number of tracers selected out of the total original samples. In this context, the term ``tracers'' denotes observed objects or regions (i.e.~terminal points, clouds, clusters, stars or masers) which allow for a measurement of the rotation curve of the Galaxy. For the sources signalled with $\dagger$, in addition to the line-of-sight velocities, we also process the measured proper motions.}\label{tab:data}
\end{center}
\end{table*}

\section{Compilation}\label{sec:comp}
\par The rationale behind our literature survey is to be as exhaustive as possible, but still selective enough to put together a clean and reliable sample of kinematic tracers. For that reason, we have decided to exclude tracers with kinematic distance determination only, important asymmetric drift or large random motions. Whereas we have taken utter care in making such compilation exhaustive, we cannot exclude that some data sets may be missing; the reader compelled to add extra data sets is welcome to do so by modifying the open source code described in Sec.~\ref{sec:tool}. Our compilation is focussed on the range of Galactocentric radii $R=3-20\,$kpc and is not intended for use far outside this range. In particular, compilations dedicated to the very internal regions of the Milky Way are available elsewhere in the literature \cite{Sofue:2013kja}. However, we warn the reader that the use of kinematic tracers at $R \lesssim 2-5\,$kpc may be problematic for certain applications \cite{2015A&A...578A..14C}.

\par Note as well that kinematic tracers beyond the ones in our compilation exist and are discussed at length in the literature, including tracers in the outer halo \cite{2006MNRAS.369.1688D,2008ApJ...684.1143X,2012ApJ...761...98K,Bhattacharjee2014,2014ApJ...794...59K}, disc \cite{Bovy2012,Bovy:2013raa,2014ApJ...794..151L,2003A&A...397..133R} and off the Galactic plane \cite{1991ApJ...367L...9K,2004MNRAS.352..440H,2012ApJ...751...30M,2012ApJ...756...89B,2015A&A...573A..91M}. For careful analyses of the implications of these and other tracers on the mass distribution of the Milky Way, we refer the reader to the works cited above and also to Refs.~\cite{2011MNRAS.416.2318G,2012ApJ...746..181S, 2012MNRAS.425.1445G, 2013ApJ...772..108Z, Read2014,DehnenBinney1998, 2009PASJ...61..227S, CatenaUllio2010, 2010A&A...509A..25W, 2010A&A...523A..83S, 2011JCAP...11..029I, 2011MNRAS.414.2446M, 2013JCAP...07..016N, 2015arXiv150405368S}, where the authors have often times built their own tracer compilations. 
Partial compilations are easily reproducible with the \texttt{galkin} tool given the possibility to (de-)select individual references. 
Again, we remind the interested reader that our code \texttt{galkin} is modular and easily modifiable by the user for the addition of new data sets.

\par We divide up the data in three main categories:
\begin{enumerate}\itemsep1pt \parskip0pt \parsep0pt
\item \emph{gas kinematics}, including
\begin{itemize}\itemsep1pt \parskip0pt \parsep0pt
\item HI terminal velocities \cite{Fich1989,Malhotra1995,McClure-GriffithsDickey2007},
\item HI thickness \cite{HonmaSofue1997},
\item CO terminal velocities \cite{BurtonGordon1978,Clemens1985,Knapp1985,Luna2006},
\item HII regions \cite{Blitz1979,Fich1989,TurbideMoffat1993,BrandBlitz1993,Hou2009}, and
\item giant molecular clouds \cite{Hou2009};
\end{itemize}

\item \emph{star kinematics}, including
\begin{itemize}\itemsep1pt \parskip0pt \parsep0pt
\item open clusters \cite{FrinchaboyMajewski2008},
\item planetary nebulae \cite{Durand1998},
\item classical cepheids \cite{Pont1994,Pont1997}, and
\item carbon stars \cite{DemersBattinelli2007,Battinelli2013}; and
\end{itemize}

\item \emph{masers} \cite{Reid2014,Honma2012,StepanishchevBobylev2011,Xu2013,BobylevBajkova2013}.
\end{enumerate}
Tab.~\ref{tab:data} recaps the key features of each data set. We refer to the original source references and in particular to the supplementary information of Ref.~\cite{2015NatPh..11..245I} for in-depth descriptions of the different object types. In order to obtain a clean sample, we have imposed various data selection cuts on the available sources following closely the recommendations of each original reference (notice, however, that the interested user can easily override these software cuts hacking the source code directly). The Appendix at the end of the present paper gives a full account of our data selection and treatment for each reference listed in Tab.~\ref{tab:data}.

\begin{figure*}[htp]
\centering
\includegraphics[width=0.325\textwidth]{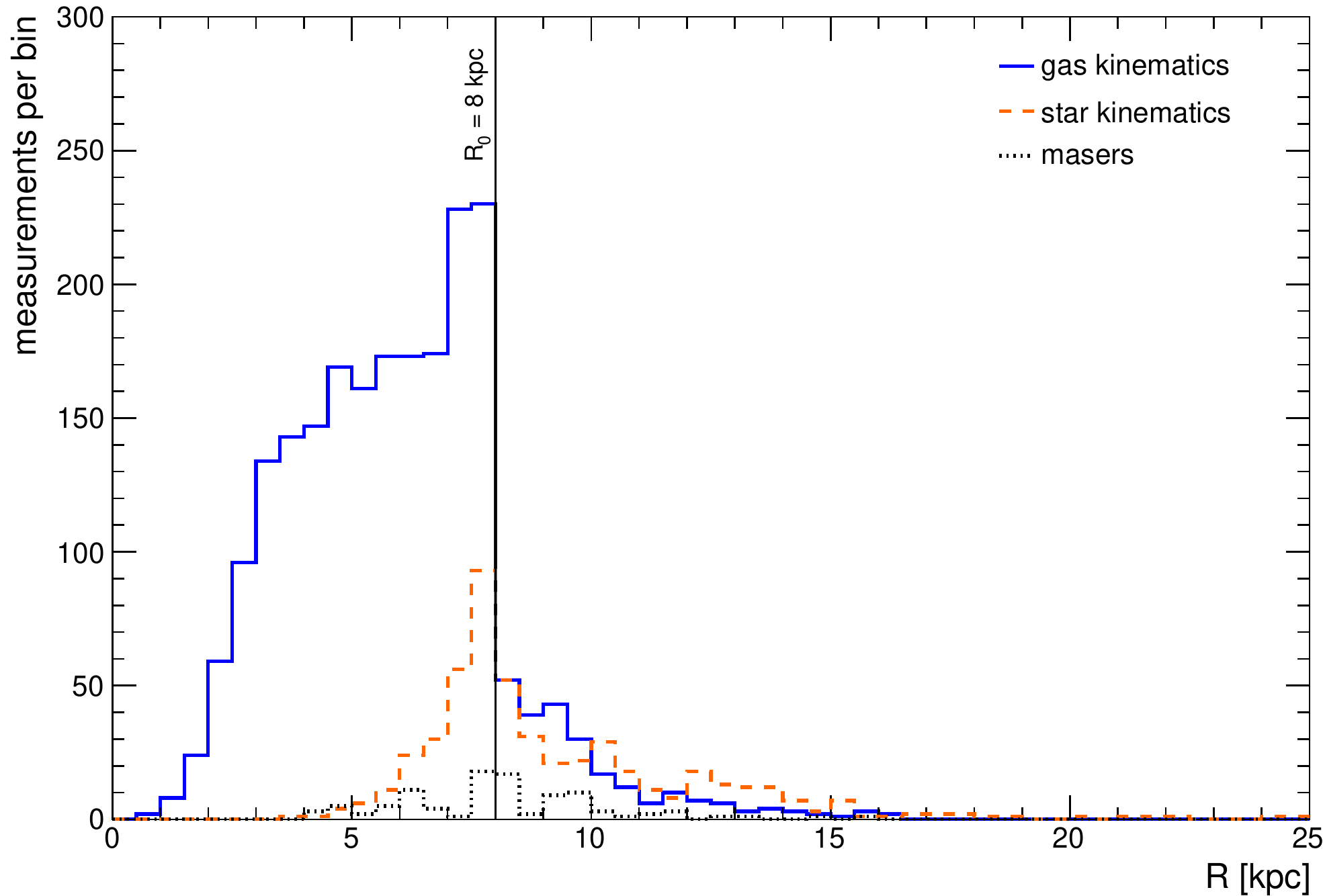}
\includegraphics[width=0.325\textwidth]{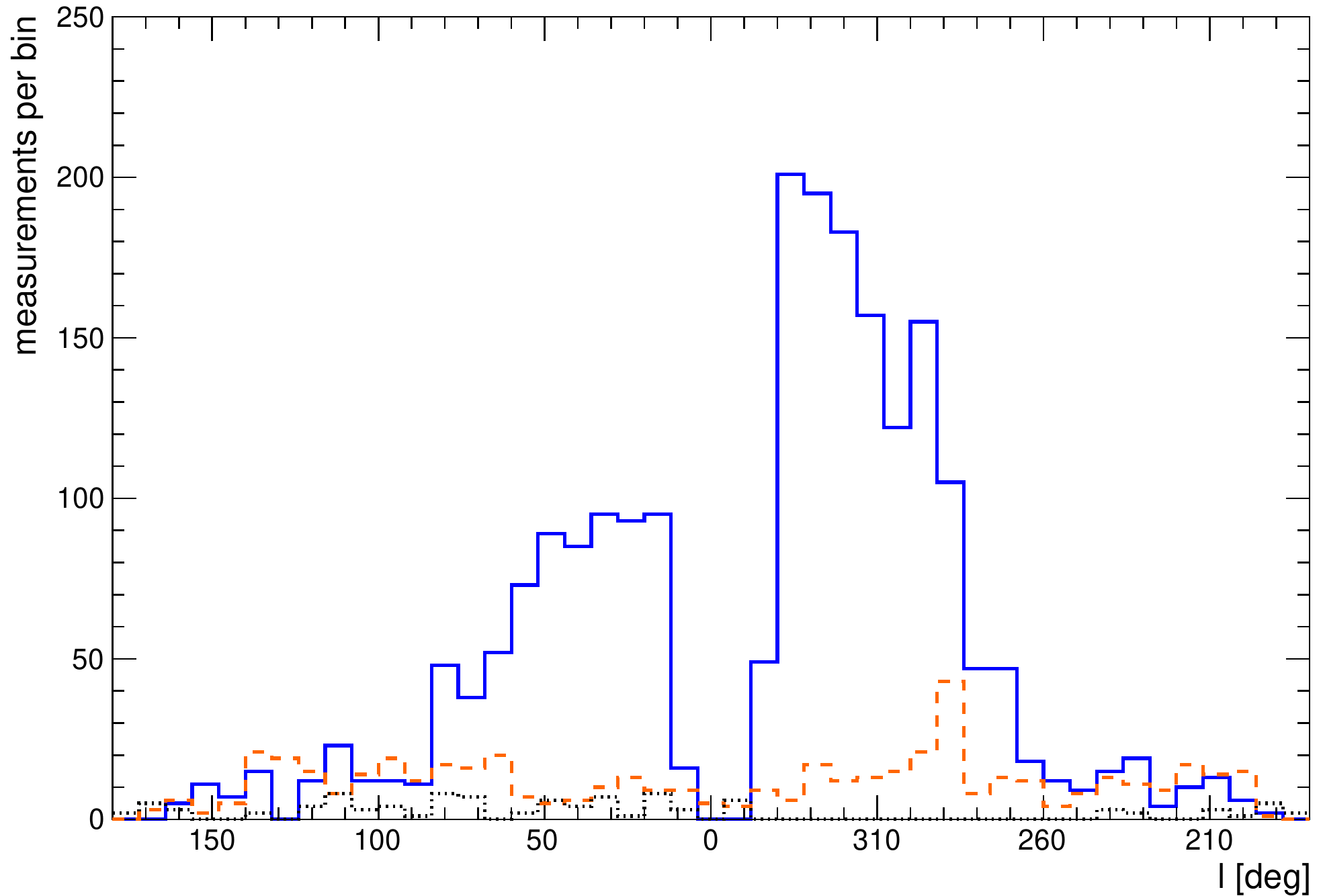}
\includegraphics[width=0.325\textwidth]{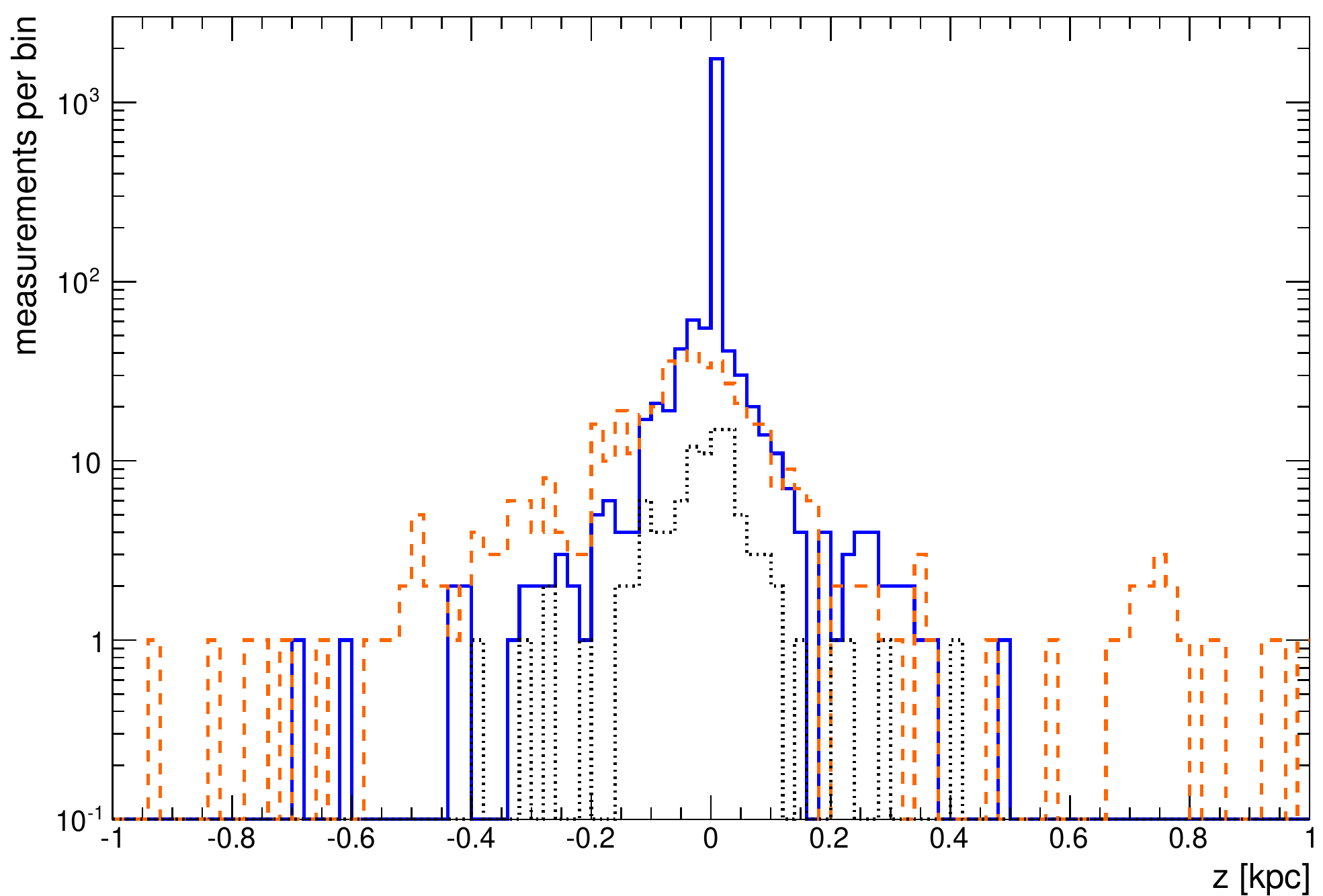}

\caption{The distribution of all kinematic tracers of the compilation in Galactocentric radius (left), Galactic longitude (centre) and height above Galactic plane (right). In each panel the blue solid, orange dashed and black dotted lines correspond to gas kinematics, star kinematics and masers, respectively. The leftmost distribution is obtained for $R_0=8\,$kpc.}
\label{fig:histos}  
\end{figure*}

\begin{figure*}[htp]
\centering
\includegraphics[width=0.495\textwidth]{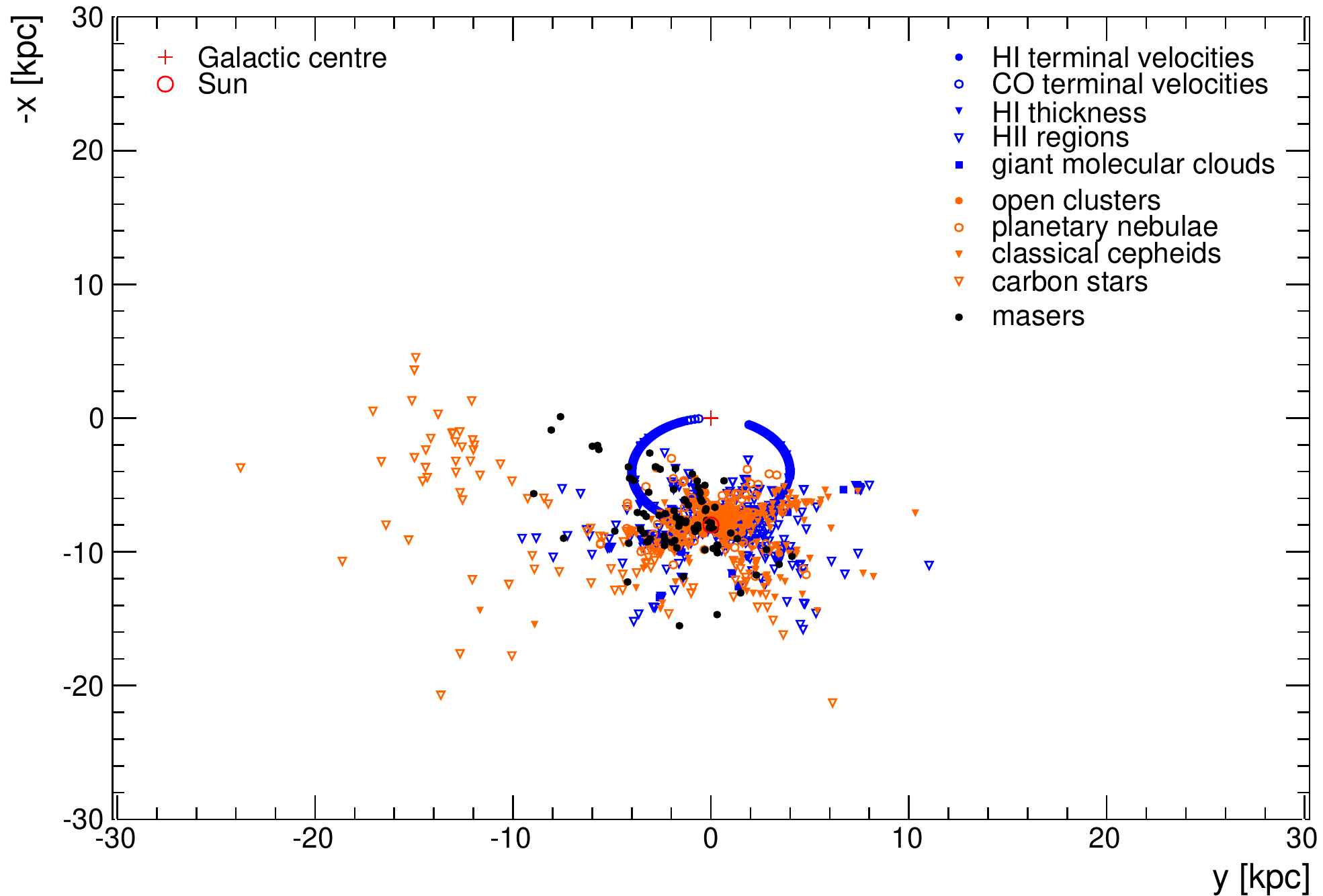}
\includegraphics[width=0.495\textwidth]{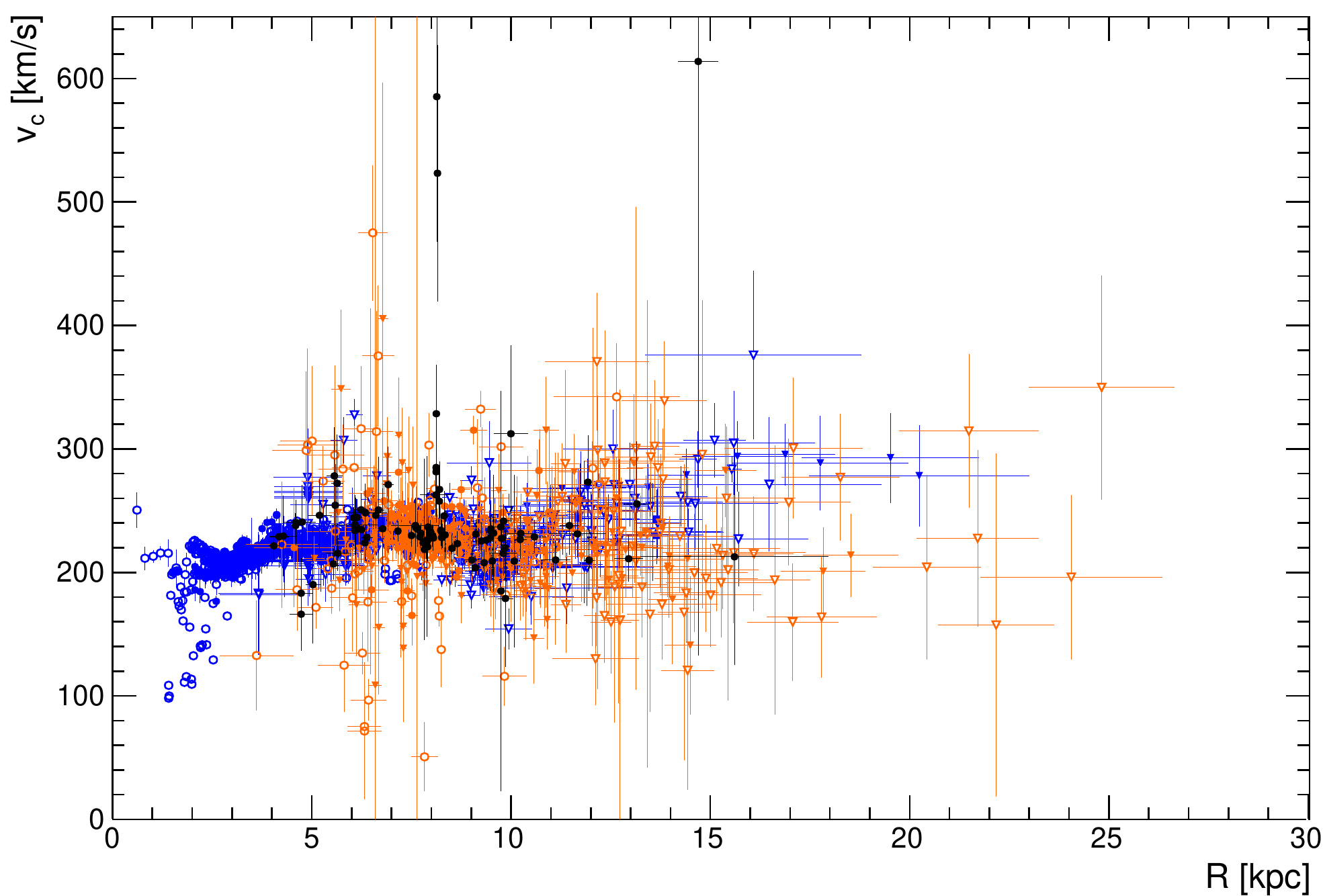}
\caption{The rotation curve of the Milky Way as derived from gas kinematics (blue), star kinematics (orange) and masers (black). The left panel shows the positions of the different tracers in the Galactic plane assuming $R_0=8\,$kpc. The Galactic centre sits at $(x,y)=(0,0)\,$kpc, while the Sun position is $(x,y)=(8,0)\,$kpc. The right panel displays the circular velocities of the tracers as a function of the Galactocentric radius assuming $R_0=8\,$kpc, $v_0=230\,$km/s and $(U,V,W)_{\odot}=(11.10,12.24,7.25)\,$km/s \cite{Schoenrich2010}. The full data displayed here is accessible via the \texttt{galkin} tool available through our GitHub page \href{https://github.com/galkintool/galkin}{\tt github.com/galkintool/galkin}.}
\label{fig:posrot}  
\end{figure*}

\par Our compilation consists of 2780 individual tracers distributed in Galactocentric radius $R$, Galactic longitude $\ell$ and height $z$ above Galactic plane as shown in Fig.~\ref{fig:histos}. Each object is specified by its coordinates $(\ell,b)$, heliocentric distance $d$ and heliocentric line-of-sight velocity $v_{\textrm{h}}^{\textrm{los}}$. The uncertainties on $\ell$ and $b$ are largely unimportant and hence neglected, whereas the uncertainties on $d$ and $v_{\textrm{h}}^{\textrm{los}}$ are taken from the original references (cf.~details in the Appendix). In radio observations, it is customary to report measurements of $v_{\textrm{h}}^{\textrm{los}}$ in terms of the line-of-sight velocity in the local standard of rest (LSR) $v_{\textrm{lsr}}^{\textrm{los}}$ for a fixed peculiar solar motion $(U,V,W)_{\odot}$. In these cases, we infer $v_{\textrm{h}}^{\textrm{los}}$ by subtracting the peculiar solar motion used in the reference off the reported $v_{\textrm{lsr,0}}^{\textrm{los}}$ (cf.~the Appendix). Once $v_{\textrm{h}}^{\textrm{los}}$ is obtained, this is summed to the adopted peculiar solar motion to get the final LSR line-of-sight velocity $v_{\textrm{lsr}}^{\textrm{los}}$. Each object has an associated measurement $(\ell,b,d\pm\Delta d,v_{\textrm{lsr}}^{\textrm{los}} \pm \Delta v_{\textrm{lsr}}^{\textrm{los}})$. The corresponding Galactocentric radius follows from simple geometry as
\begin{equation}\label{eq:R}
R=(d^2 \cos^2 b + R_0^2 - 2 R_0 d \cos b \, \cos \ell)^{1/2} \, ,
\end{equation}
where $R_0$ is the distance of the Sun to the Galactic centre. Under the assumption of circular orbits, the angular circular velocity of the object $\omega_c$ is found by inverting
\begin{equation}\label{eq:vlos}
v_{\textrm{lsr}}^{\textrm{los}} = \left( R_0\omega_c - v_0  \right) \cos b \, \sin \ell \, ,
\end{equation}
where $v_0$ is the local circular velocity. The uncertainties on $d$ and $v_{\textrm{lsr}}^{\textrm{los}}$ are propagated to $R$ and $\omega_c$, respectively. We shall also provide the familiar circular velocity $v_c\equiv R\omega_c$ and corresponding uncertainties, but note that the errors of $R$ and $v_c$ are strongly positively correlated, while those of $R$ and $\omega_c$ are independent.
All uncertainties currently implemented in \texttt{galkin} are symmetric following the information available in each reference; future data might provide the full distribution of observables, which would then be treated in upcoming versions of \texttt{galkin} and would be of great value for Bayesian studies. 
The procedure described above is common to all object types in Tab.~\ref{tab:data}, with some modifications in two cases. For terminal velocities, we set $b=0$ and $R=R_0|\sin\ell|$ (or, equivalently, $d=R_0|\cos\ell|$) in Eqs.~\eqref{eq:R} and \eqref{eq:vlos}, and each measurement reads $(\ell,v_{\textrm{lsr}}^{\textrm{los}} \pm \Delta v_{\textrm{lsr}}^{\textrm{los}})$. For the HI thickness method, the measured quantity is $W\equiv R_0\omega_c -v_0$ instead of $v_{\textrm{lsr}}^{\textrm{los}}$, so each data point is defined by $(R/R_0 \pm \Delta R/R_0, W \pm \Delta W)$, cf.~Refs.~\cite{1992AJ....103.1552M,HonmaSofue1997}. We also process the proper motions $\mu_{\ell^\ast},\mu_b$ when available, as is the case of open clusters and masers. All other details of data treatment are exhaustively documented in the Appendix.

\par Note that Eqs.~\eqref{eq:R} and \eqref{eq:vlos} can be used to infer $(R,\omega_c)$ only upon fixing of $R_0$, $v_0$ and $(U,V,W)_{\odot}$, which are quantities still affected by sizeable uncertainties -- see Refs.~\cite{Gillessen2009,Ando2011,Malkin2012,Reid2014} for $R_0$, Refs.~\cite{ReidBrunthaler2004,Reid2009,Bovy2009, McMillan2010, Bovy2012,Reid2014} for $v_0$ and Refs.~\cite{DehnenBinney1998,Schoenrich2010,Bovy2012,Reid2014} for $(U,V,W)_{\odot}$. Making the reasonable choice of Galactic parameters $R_0=8\,$kpc, $v_0=230\,$km/s and $(U,V,W)_{\odot}=(11.10,12.24,7.25)\,$km/s \cite{Schoenrich2010}, we show in Fig.~\ref{fig:posrot} the positions and circular velocities of all tracers implemented. This is the main output of \texttt{galkin}.

\section{Tool description}\label{sec:tool}
\par We now turn to the description of the \texttt{galkin} tool, whose function is to allow the user to access the data described Sec.~\ref{sec:comp} in a customisable way. The tool is written in Python and the code package is available through our GitHub page, \href{https://github.com/galkintool/galkin}{\tt github.com/galkintool/galkin}.

\par The distribution has a very simple structure. The parent directory contains the setup file \texttt{setup.py} together with the usual \texttt{README} file. The \texttt{galkin} package is provided under \texttt{galkin/}, where \texttt{galkin/data/} contains the original data as extracted from the corresponding references\footnote{Note that the main point of this tool is to handle in a unified way data from references using different Galactic parameters, different definitions and often overlapping in observed sources. The data files under \texttt{data/} contain the original published data of each reference; please do not use these files directly unless you are fully aware of all details of each reference.}, and example scripts can be found under \texttt{bin/}. The main dependencies and installation instructions are described in the \texttt{README} file, whereas here we simply summarise the usage of the tool assuming it is properly installed and running.
Note that \texttt{galkin} adopts \texttt{astropy} \cite{2013A&A...558A..33A} for coordinate conversion if required by the user, but the code can also be used without installing this package, which is the default option.

\par
\begin{figure}[tp]
\centering
\includegraphics[width=0.4\textwidth]{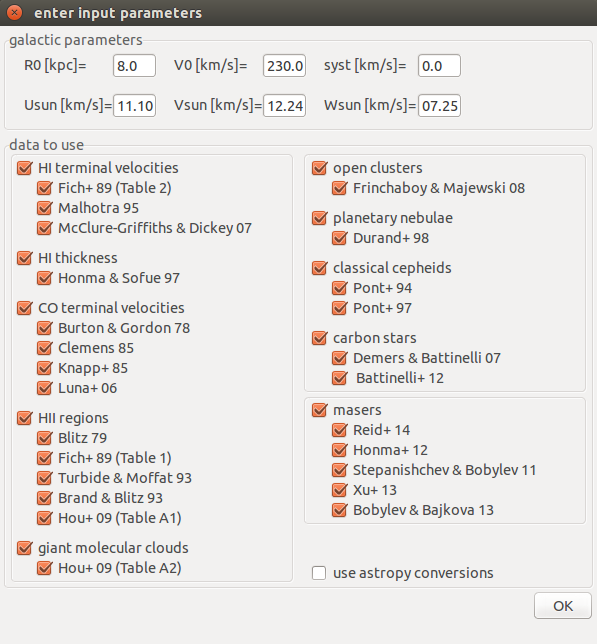}
\caption{The graphic interface window of \texttt{galkin} in Ubuntu 14.04.}
\label{fig:interface}  
\end{figure}

 The goal of \texttt{galkin} is to provide ready-to-use data files containing all the necessary kinematic tracer information for constraining the rotation curve of the Galaxy. The user may choose the values for the Galactic parameters $(R_0,v_0)$ and $(U,V,W)_{\odot}$, as well as (de)select either entire classes of tracers (gas, stars, masers) or single references in Tab.~\ref{tab:data}. It is also possible to add a given systematic uncertainty to the line-of-sight velocity of each tracer. This is done with the help of the script \texttt{bin/galkin\_data.py} either through a graphic interface (see Fig.~\ref{fig:interface}) by launching the tool from \texttt{bin/} with the command
\vspace{0.1cm}
\par \texttt{python galkin\_data.py}
\vspace{0.15cm}
\par \hspace{-0.5cm} or through a customisable input file by typing
\vspace{0.1cm}
\par \texttt{python galkin\_data.py inputpars.txt}
\vspace{0.15cm}
\par \hspace{-0.5cm} The code then processes the original data sets of the selected references converting each data set consistently for the chosen values of the Galactic parameters. The output is stored in three data files:
\vspace{-0.2cm}
\begin{itemize}[leftmargin=*] \itemsep0pt \parskip0pt \parsep0pt
\item[] \texttt{bin/output/posdata.dat} with the position information of each tracer, namely $(R,d,\ell,b,x,y,z)$;
\item[] \texttt{bin/output/vcdata.dat} with the rotation curve measurements, namely $(R,\Delta R,v_c,\Delta v_c,\omega_c,\Delta \omega_c)$; and
\item[] \texttt{bin/output/rawdata.dat} with the raw data measurements, namely $(v_\text{los},\Delta v_\text{los}, \mu_{\ell^\ast},\Delta \mu_{\ell^\ast}, \mu_b, \Delta \mu_b)$.
\end{itemize}
\vspace{-0.2cm}
In the first two files, the values of the Galactic parameters chosen by the user are reported in the first line and the source reference for each tracer is indicated in the last column. For testing purposes, we provide under \texttt{bin/output/} the sample files corresponding to our baseline choice of Galactic parameters used to produce Fig.~\ref{fig:posrot} (for the entire database and single classes objects, separately) as well as for a variation of the Galactic parameters (for the entire database only).

\par The tool also includes the script \texttt{bin/galkin\_plotter.py} to read and visualise the output described above. This can be launched from \texttt{bin/} with the command
\vspace{0.1cm}
\par \texttt{python galkin\_plotter.py output/vcdata.dat output/posdata.dat}
\vspace{0.15cm}
\par \hspace{-0.5cm}
which produces a set of demonstrative plots including the spatial distribution of the tracers and the inferred rotation curve.

Finally, we provide the script \texttt{bin/galkin\_data\_fast.py} in order to illustrate how to use \texttt{galkin} inside another code without dealing with input nor output files. This is a faster version of the data processing pipeline that is specifically designed for applications that need to use \texttt{galkin} repeateadly, e.g.~in scans over Galactic parameters.

\par The structure of \texttt{galkin} is purposely minimal and modular. The code can be easily adapted to replace or modify existing data sets or single tracers and to add new data sets as they become available. The idea behind \texttt{galkin} is to provide an up-to-date and user-friendly compilation of tracers of the kinematics of the Milky Way over the coming years.


\vspace{0.5cm}
{\it Acknowledgements.} M.~P.~acknowledges the support from Wenner-Gren Stiftelserna in Stockholm; F.~I.~from the Simons Foundation and FAPESP process 2014/11070-2.

\appendix*

\section{Data treatment}
\par Tracers of the rotation curve of the Milky Way usually adopted in the literature can be roughly divided into three categories: gas kinematics, kinematics of stellar objects, and masers. For each of these classes of objects, different methodologies are used to infer positions, distances and kinematics. Without attempting here a review of the properties of each tracer type, we point to the supplementary information of Ref.~\cite{2015NatPh..11..245I}, where this collection of data has first been presented. The reader will find there an in-depth description of all tracer types as well as appropriate references to the original literature. In this Appendix, we recap the original source references and fully document our data treatment. Let us notice that halo stars are currently not included in this version of \texttt{galkin}, due to the additional assumptions needed. Future versions of the code will properly include halo stars as an extra tracer type.

\par Tab.~\ref{tab:data} displays the key features of all data sets used in \texttt{galkin}. In the following we provide a detailed description of the treatment applied to each data set. Equation, figure and table numbering refers to the original source references.

\subsection{Gas kinematics}

\subsubsection{HI terminal velocities}

\subparagraph{Fich+ '89 \cite{Fich1989}} From Tab.~2, we take the terminal velocities $v_{\textrm{lsr,0}}^{\textrm{los}}$ measured at different longitudes $\ell$ and assume an overall velocity uncertainty $\Delta v_{\textrm{lsr}}^{\textrm{los}}=4.5\,$km/s following Sec.~II.b.i. The reported velocities $v_{\textrm{lsr,0}}^{\textrm{los}}$ correspond to a peculiar solar motion $(U,V,W)_{\odot}=(10.3,15.3,7.7)\,$km/s, which is consistently subtracted off in our compilation\footnote{\label{LSRfootnote} Notice that the original reference fails to indicate explicitely the adopted peculiar solar motion. Following Ref.~\cite{Reid2009}, we assume the adopted value is the old standard solar motion as defined in the Appendix of Ref.~\cite{Reid2009}.}. We further correct for the peculiar LSR motion of $4.2\,$km/s in the radial direction, cf.~Sec.~II.b.i and Eq.~(4)\footnote{\label{correctionfootnote} Notice that this correction depends on the assumed peculiar solar motion, so this procedure is strictly valid only when the user selects the same peculiar solar motion as in the original reference; in practice, however, this is a small correction and we ignore the slight inconsistency.}.

\subparagraph{Malhotra '95 \cite{Malhotra1995}} From Fig.~7, we take the terminal velocities $v_{\textrm{lsr,0}}^{\textrm{los}}$ measured at different Galactocentric radii $R/R_0$ and assume an overall velocity uncertainty $\Delta v_{\textrm{lsr}}^{\textrm{los}}=9\,$km/s in line with the dispersions computed in Sec.~3.4 for both the first and fourth quadrants. The Galatocentric radii are converted into longitudes through $R=R_0|\sin\ell|$ depending on the quadrant (first quadrant: circles and triangles in top panel of Fig.~7; fourth quandrant: squares in bottom panel of Fig.~7). The reported velocities $v_{\textrm{lsr,0}}^{\textrm{los}}$ correspond to a peculiar solar motion $(U,V,W)_{\odot}=(10.3,15.3,7.7)\,$km/s, which is consistently subtracted off in our compilation$^\text{\ref{LSRfootnote}}$.

\subparagraph{McClure-Griffiths \& Dickey '07 \cite{McClure-GriffithsDickey2007}} From Tab.~1 (online version), we take the terminal velocities $v_{\textrm{lsr,0}}^{\textrm{los}}$ measured at different longitudes $\ell$. According to Secs.~3.3 and 4.3.1 and the caption of Fig.~8, the velocity uncertainty amounts to $\Delta v_{\textrm{lsr}}^{\textrm{los}}=1\,$km/s for $\ell<325^{\circ}$, $\Delta v_{\textrm{lsr}}^{\textrm{los}}=3\,$km/s for $\ell>332.5^{\circ}$ and $\Delta v_{\textrm{lsr}}^{\textrm{los}}=10\,$km/s for $\ell=327.5^{\circ}-332^{\circ}$; we conservatively assume $\Delta v_{\textrm{lsr}}^{\textrm{los}}=10\,$km/s for the whole range $\ell=325^{\circ}-332.5^{\circ}$. The reported velocities $v_{\textrm{lsr,0}}^{\textrm{los}}$ correspond to a peculiar solar motion $(U,V,W)_{\odot}=(10.3,15.3,7.7)\,$km/s, which is consistently subtracted off in our compilation$^\text{\ref{LSRfootnote}}$. We exclude from the full data set regions with discrete HI clouds at $\ell=306^{\circ}\pm 1^{\circ},312^{\circ}\pm 0.5^{\circ},320^{\circ}\pm 0.5^{\circ}$ (cf.~Sec.~4; the width of the intervals is fixed a posteriori in order to eliminate the spikes in velocity) and also the region at $|\sin\ell|>0.95$ because there $\Delta v_{\textrm{lsr}}^{\textrm{los}} / v_{\textrm{lsr,0}}^{\textrm{los}}\sim 1$ (cf.~Sec.~4.2; this last cut is actually already performed in the online version of Tab.~1).

\subsubsection{HI thickness method}

\subparagraph{Honma \& Sofue '97 \cite{HonmaSofue1997}} From Tab.~1, we take the Galactocentric radii $R/R_0$ fitted through the thickness method for different velocities $W\equiv R_0\omega_c -v_0$ and assume an overall uncertainty $\Delta W=5.8\,$km/s following Sec.~2.4. The authors of this reference find that the method of Merrifield '92 \cite{1992AJ....103.1552M} (i.e.~method 1 in Tab.~1) covers the largest range of $R/R_0$ and is the most accurate, so we use the results of that method for our compilation.

\subsubsection{CO terminal velocities}

\subparagraph{Burton \& Gordon '78 \cite{BurtonGordon1978}} From Fig.~2, we take the terminal velocities $v_{\textrm{lsr,0}}^{\textrm{los}}$ measured at different longitudes $\ell$, where $v_{\textrm{lsr,0}}^{\textrm{los}}$ already includes the line width correction, cf.~Sec.~3 and Eq.~(4a). The resolution of the original CO data presented in this reference is $\Delta v_{\textrm{lsr}}^{\textrm{los}}=1.3\,$km/s, while it amounts to $\Delta v_{\textrm{lsr}}^{\textrm{los}}=2.6\,$km/s for the other data sets \cite{Bania1977,Liszt1977} plotted in Fig.~2 (cf.~Sec.~2); therefore, we conservatively assume $\Delta v_{\textrm{lsr}}^{\textrm{los}}=2.6\,$km/s all along. The reported velocities $v_{\textrm{lsr,0}}^{\textrm{los}}$ correspond to a peculiar solar motion $(U,V,W)_{\odot}=(10.3,15.3,7.7)\,$km/s, which is consistently subtracted off in our compilation.

\subparagraph{Clemens '85 \cite{Clemens1985}} From Tab.~2, we take the terminal velocities $\tilde{v}_{\textrm{lsr,0}}^{\textrm{los}}$ measured at different longitudes $\ell$ and the corresponding uncertainties $\Delta \tilde{v}_{\textrm{lsr}}^{\textrm{los}}$, which actually represent 0.67 of the standard deviation, cf.~footnote c in Tab.~2. The terminal velocities $\tilde{v}_{\textrm{lsr,0}}^{\textrm{los}}$ in Tab.~2 are uncorrected for the line width, so the true terminal velocities read $v_{\textrm{lsr,0}}^{\textrm{los}}=\tilde{v}_{\textrm{lsr,0}}^{\textrm{los}}-3\,$km/s (cf.~Sec.~II.c) and the corresponding uncertainties are $\Delta v_{\textrm{lsr}}^{\textrm{los}}=\left( (\Delta \tilde{v}_{\textrm{lsr}}^{\textrm{los}}/0.67)^2 +(0.6\,\textrm{km/s})^2  \right)^{1/2}$, where $0.6\,$km/s is the typical error in the line width correction, cf.~Sec.~II.c. The reported velocities $v_{\textrm{lsr,0}}^{\textrm{los}}$ correspond to a peculiar solar motion $(U,V,W)_{\odot}=(10.3,15.3,7.7)\,$km/s, which is consistently subtracted off in our compilation$^\text{\ref{LSRfootnote}}$. We further correct for the peculiar LSR motion of $7\,$km/s in the azimuthal direction, cf.~Sec.~II.d$^\text{\ref{correctionfootnote}}$.

\subparagraph{Knapp+ '85 \cite{Knapp1985}} From Fig.~5, we take the terminal velocities $v_{\textrm{lsr,0}}^{\textrm{los}}\pm \Delta \tilde{v}_{\textrm{lsr}}^{\textrm{los}}$ (top panel) and the velocity dispersions $\sigma$ (central panel) measured at different Galactocentric radii $R/R_0$. The statistical uncertainty of the terminal velocities is summed in quadrature to the dispersion, i.e.~$\Delta v_{\textrm{lsr}}^{\textrm{los}}=\left( (\Delta \tilde{v}_{\textrm{lsr}}^{\textrm{los}})^2 + \sigma^2 \right)^{1/2}$. The Galatocentric radii are converted into longitudes through $R=R_0|\sin\ell|$ considering that the data refers to the first quadrant only. The reported velocities $v_{\textrm{lsr,0}}^{\textrm{los}}$ correspond to a peculiar solar motion $(U,V,W)_{\odot}=(10.3,15.3,7.7)\,$km/s, which is consistently subtracted off in our compilation$^\text{\ref{LSRfootnote}}$.

\subparagraph{Luna+ '06 \cite{Luna2006}} From Fig.~1 (top panel), we take the terminal velocities $v_{\textrm{lsr,0}}^{\textrm{los}}$ measured at different longitudes $\ell$ and assume an overall velocity uncertainty $\Delta v_{\textrm{lsr}}^{\textrm{los}}=3\,$km/s following Sec.~3.1. The reported velocities $v_{\textrm{lsr,0}}^{\textrm{los}}$ correspond to a peculiar solar motion $(U,V,W)_{\odot}=(14.8/20.)\times(10.3,15.3,7.7)\,$km/s, which is consistently subtracted off in our compilation$^\text{\ref{LSRfootnote}}$. We exclude from the full data set the region at $\ell=280^{\circ}-312^{\circ}$ due to the influence of the Carina arm and Centaurus, cf.~Sec.~III.a.iii in Ref.~\cite{Alvarez1990}.

\subsubsection{HII regions}

\subparagraph{Blitz '79 \cite{Blitz1979}} From Tab.~2, we take the distances $d\pm \Delta d$ and line-of-sight velocities $v_{\textrm{lsr,0}}^{\textrm{los}}\pm \Delta v_{\textrm{lsr}}^{\textrm{los}}$ corresponding to objects towards different Galactic coordinates $(\ell,b)$. The reported velocities $v_{\textrm{lsr,0}}^{\textrm{los}}$ correspond to a peculiar solar motion $(U,V,W)_{\odot}=(10.3,15.3,7.7)\,$km/s, which is consistently subtracted off in our compilation$^\text{\ref{LSRfootnote}}$. We exclude from the full data set all objects in (or coincident to objects in) Fich+ '89, Turbide \& Moffat '93, Brand \& Blitz '93 and Hou+ '09; only three objects remain: S148, S306 and Mon R2.

\subparagraph{Fich+ '89 \cite{Fich1989}} From Tab.~1, we take the distances $d\pm \Delta d$ and line-of-sight velocities $v_{\textrm{lsr,0}}^{\textrm{los}}\pm \Delta \tilde{v}_{\textrm{lsr}}^{\textrm{los}}$ corresponding to objects towards different Galactic coordinates $(\ell,b)$ and add a random motion of $6.4\,$km/s in quadrature to $\Delta \tilde{v}_{\textrm{lsr}}^{\textrm{los}}$ following Sec.~II.i.b. The reported velocities $v_{\textrm{lsr,0}}^{\textrm{los}}$ correspond to a peculiar solar motion $(U,V,W)_{\odot}=(10.3,15.3,7.7)\,$km/s, which is consistently subtracted off in our compilation$^\text{\ref{LSRfootnote}}$. We further correct for the peculiar LSR motion of $4.2\,$km/s in the radial direction, cf.~Sec.~II.b.i and Eq.~(4)$^\text{\ref{correctionfootnote}}$. We exclude from the full data set all objects in (or coincident to objects in) Turbide \& Moffat '93 and Brand \& Blitz '93.

\subparagraph{Turbide \& Moffat '93 \cite{TurbideMoffat1993}} From Tab.~5 (for $Z=Z(R)$), we take the distances $d\pm \Delta d$ and line-of-sight velocities $v_{\textrm{lsr,0}}^{\textrm{los}}\pm \Delta v_{\textrm{lsr}}^{\textrm{los}}$ corresponding to objects towards different Galactic coordinates $(\ell,b)$. Note that the reported velocities $v_{\textrm{lsr,0}}^{\textrm{los}}$ correspond to Eq.~(11) with $V_{\text{mol}}=0\,$km/s and a peculiar solar motion $(U,V,W)_{\odot}=(9,11,6)\,$km/s, which is consistently subtracted off in our compilation. For the objects in Tab.~5 based on Refs.~[1,3] therein, we further correct for the peculiar LSR motion of $4.2\,$km/s in the radial direction, cf.~Sec.~4 and Eq.~(11). We exclude from the full data set all objects in (or coincident to objects in) Brand \& Blitz '93.

\subparagraph{Brand \& Blitz '93 \cite{BrandBlitz1993}} From Tab.~1, we take the distances $d\pm \Delta d$ and line-of-sight velocities $v_{\textrm{lsr,0}}^{\textrm{los}}\pm \Delta \tilde{v}_{\textrm{lsr}}^{\textrm{los}}$ corresponding to objects towards different Galactic coordinates $(\ell,b)$ and add a random motion of $6.4\,$km/s in quadrature to $\Delta \tilde{v}_{\textrm{lsr}}^{\textrm{los}}$ following Sec.~3.1. The reported velocities $v_{\textrm{lsr,0}}^{\textrm{los}}$ correspond to a peculiar solar motion $(U,V,W)_{\odot}=(10.3,15.3,7.7)\,$km/s, which is consistently subtracted off in our compilation$^\text{\ref{LSRfootnote}}$. We exclude from the full data set objects near to the Galactic centre $\ell=345^{\circ}-15^{\circ}$ or anti Galactic centre $\ell=165^{\circ}-195^{\circ}$ (cf.~Sec.~3.3) and nearby objects at $d<1\,$kpc (cf.~Sec.~3.3).

\subparagraph{Hou+ '09 \cite{Hou2009}} From Tab.~A1, we take the distances $d\pm \Delta d$ and line-of-sight velocities $v_{\textrm{lsr,0}}^{\textrm{los}}$ corresponding to objects towards different Galactic coordinates $(\ell,b)$ and assume an overall velocity uncertainty $\Delta v_{\textrm{lsr}}^{\textrm{los}}=3\,$km/s. The reported velocities $v_{\textrm{lsr,0}}^{\textrm{los}}$ correspond to a peculiar solar motion $(U,V,W)_{\odot}=(10.3,15.3,7.7)\,$km/s, which is consistently subtracted off in our compilation$^\text{\ref{LSRfootnote}}$. We exclude from the full data set objects without stellar distances nor at the tangent points, near to the Galactic centre $\ell=345^{\circ}-15^{\circ}$ or anti Galactic centre $\ell=165^{\circ}-195^{\circ}$ and objects in (or coincident to objects in) Brand \& Blitz '93.

\subsubsection{Giant molecular clouds}

\subparagraph{Hou+ '09 \cite{Hou2009}} From Tab.~A2, we take the distances $d$ and line-of-sight velocities $v_{\textrm{lsr,0}}^{\textrm{los}}$ corresponding to objects towards different Galactic coordinates $(\ell,b)$ and adopt an overall velocity uncertainty $\Delta v_{\textrm{lsr}}^{\textrm{los}}=3\,$km/s and an overall relative distance uncertainty $\Delta d/d=20\%$. The reported velocities $v_{\textrm{lsr,0}}^{\textrm{los}}$ correspond to a peculiar solar motion $(U,V,W)_{\odot}=(10.3,15.3,7.7)\,$km/s, which is consistently subtracted off in our compilation$^\text{\ref{LSRfootnote}}$. We exclude from the full data set objects without stellar distances and near to the Galactic centre $\ell=345^{\circ}-15^{\circ}$ or anti Galactic centre $\ell=165^{\circ}-195^{\circ}$.

\subsection{Star kinematics}

\subsubsection{Open clusters}

\subparagraph{Frinchaboy \& Majewski '08 \cite{FrinchaboyMajewski2008}} From Tab.~1, we take the distances $d$ corresponding to objects towards different equatorial coordinates $(\alpha,\delta)$ (then converted to Galactic coordinates $(\ell,b)$) based on Ref.~\cite{Dias2002} and assume an overall relative distance uncertainty $\Delta d/d=20\%$. From Tab.~12, we take the heliocentric line-of-sight velocities $v_{\textrm{h}}^{\textrm{los}}\pm \Delta v_{\textrm{h}}^{\textrm{los}}$, where whenever possible we use the bulk kinematics derived with the three-dimensional and radial velocity membership criterion (3D+RV), and also the proper motions $\mu_{\ell^\ast} \pm \Delta \mu_{\ell^\ast}$, $\mu_b \pm \Delta \mu_b$. We exclude from the full data set objects near to the anti Galactic centre $\ell=160^{\circ}-200^{\circ}$; NGC 1513, NGC 7654 (only one member identified, cf.~Sec.~6.1); Collinder 258, Lynga 1, NGC 6250 (probably wrong memberships, cf.~Sec.~6.2.3); and NGC 6416 (high residual velocity).

\subsubsection{Planetary nebulae}

\subparagraph{Durand+ '98 \cite{Durand1998}} From Tab.~2 (online version), we take the heliocentric line-of-sight velocities $v_{\textrm{h}}^{\textrm{los}}\pm \Delta v_{\textrm{h}}^{\textrm{los}}$ corresponding to objects towards different equatorial coordinates $(\alpha,\delta)$ (then converted to Galactic coordinates $(\ell,b)$). The distances $d$ are found by cross-matching Tab.~2 to Tabs.~1 and 3 of Ref.~\cite{Zhang1995}, which report individually determined and statistical distances, respectively. When available, we prefer the individually determined distances (i.e.~Tab.~1 of Ref.~\cite{Zhang1995}), based on the methods presented in Ref.~\cite{Zhang1993} with a relative uncertainty ranging from 15 to 40\% (cf.~Sec.~3.3 in Ref.~\cite{Zhang1993}) so an overall relative distance uncertainty $\Delta d/d=25\%$ is adopted; when not possible, we use the statistical distance scale (i.e.~Tab.~3 of Ref.~\cite{Zhang1995}), for which the relative uncertainty is $\Delta d/d=30\%$ (cf.~Sec.~5.2 in Ref.~\cite{Zhang1995}). We further correct for the $K$-term of $5.1\,$km/s (cf.~Eq.~(4) and Tab.~3) and model the asymmetric drift with respect to the peculiar solar motion in use (i.e.~$U_{\textrm{ad}}=16.0\,\textrm{km/s}-U_\odot$ in the radial direction and $V_{\textrm{ad}}=24.8\,\textrm{km/s}-V_\odot$ in the azimuthal direction; cf.~Eq.~(4) and Tab.~3). The error of $K$, $U_{\textrm{ad}}$ and $V_{\textrm{ad}}$ (cf.~Tab.~3) are propagated to the line-of-sight velocity $v_{\textrm{lsr}}^{\textrm{los}}$. We exclude from the full data set objects near to the Galactic centre $\ell=353^{\circ}-7^{\circ}$ and off the Galactic plane $z\geq 200\,$pc (cf.~Sec.~4.2); NGC 6565, M2-29, M1-46, Sa 1-8, NGC 6741, NGC 7293, Vy 2-2, NGC 6702, NGC 7094, NGC 2392, NGC 4071, NGC 6026, NGC 6302, M2-7, Th 3-14 (deficient statistical distances, cf.~Sec.~5.2 in Ref.~\cite{Zhang1995}; notice typo in this reference where NGC 6320 is mentioned instead of NGC 6302); BoBn 1 (atypical motion; cf.~Sec.~4.1) and NGC 6567, A 8, Pu 1, M1-5 (high residual velocity).

\subsubsection{Classical cepheids}

\subparagraph{Pont+ '94 \cite{Pont1994}} From Tab.~3, we take the distance moduli $\mu$ (FW) and heliocentric line-of-sight velocities $v_{\textrm{h}}^{\textrm{los}}$ corresponding to objects towards different Galactic coordinates $(\ell,b)$ and assume an overall distance modulus uncertainty $\Delta \mu=0.23\,$mag following Sec.~11.3. The distance moduli are converted to distances through $d/\textrm{pc}=10^{\mu/5+1}$. The velocity uncertainty reads $\Delta v_{\textrm{h}}^{\textrm{los}}=(\sigma_1^2+\sigma_2^2)^{1/2}$, where $\sigma_1=1,\, 2.5,\, 5\,$km/s depending on the method used to calculate the radial velocity (cf.~Sec.~11.3) and $\sigma_2\sim11.1\,$km/s is the contribution of the velocity ellipsoid (cf.~Sec.~11.3). We further correct for the $K$-term of $-1.81\,$km/s, cf.~Sec.~11.6. We use the reduced sample of 266 stars (cf.~Sec.~11.4 and Tab.~4) excluding objects near to the anti Galactic centre $\ell=160^{\circ}-200^{\circ}$ and two further objects due to high residual velocity.

\subparagraph{Pont+ '97 \cite{Pont1997}} From Tab.~1, we take the heliocentric line-of-sight velocities $v_{\textrm{h}}^{\textrm{los}}$, period $P$ and colours $V$ and $B-V$ corresponding to objects towards different Galactic coordinates $(\ell,b)$. The distance moduli are found with the period-luminosity-colour relation described in Ref.~\cite{Pont1994} (FW) and the zero point given in Sec.~3.1; an overall distance modulus uncertainty $\Delta \mu=0.21\,$mag is assumed following Sec.~3.3. The distance moduli are converted to distances through $d/\textrm{pc}=10^{\mu/5+1}$. The velocity uncertainty reads $\Delta v_{\textrm{h}}^{\textrm{los}}=(\sigma_1^2+\sigma_2^2)^{1/2}$, where $\sigma_1=1\,(2.5)\,$km/s for $\geq10\,(<10)$ velocity measurements (cf.~Tab.~1) and $\sigma_2\sim11.1\,$km/s is the contribution of the velocity ellipsoid (cf.~Sec.~11.3 of Ref.~\cite{Pont1994}). We further add a $6\,$km/s systematic uncertainty (cf.~Sec.~6; see also Sec.~5.3) to the derived circular velocity. We exclude from the full data set objects near to the anti Galactic centre $\ell=160^{\circ}-200^{\circ}$ and with no measured radial velocity or no $B-V$.

\subsubsection{Carbon stars}

\subparagraph{Demers \& Battinelli '07 \cite{DemersBattinelli2007}} From Tab.~4 (online version), we take the heliocentric line-of-sight velocities $v_{\textrm{h}}^{\textrm{los}}\pm \Delta \tilde{v}_{\textrm{h}}^{\textrm{los}}$ corresponding to objects towards different Galactic coordinates $(\ell,b)$ and add a random motion of $20\,$km/s in quadrature to $\Delta \tilde{v}_{\textrm{h}}^{\textrm{los}}$ following Sec.~5.1. From Tab.~1 (online version), we take the distances $d$ and assume an overall relative distance uncertainty $\Delta d/d=10\%$ following Sec.~5.1. We exclude from the full data set objects near to the anti Galactic centre $\ell=170^{\circ}-190^{\circ}$; stars no.~20 and 23 (nearby fast stars; cf.~Sec.~5.2) and stars no.~17, 18, 27, 28, 35, 42, 52, 56, 58 and 60 (possibly belonging to Canis Major; cf.~Sec.~5.2).

\subparagraph{Battinelli+ '13 \cite{Battinelli2013}} From Tab.~1, we take the Galactocentric radii $\tilde{R}$ (computed with $R_0=7.62\,$kpc) and heliocentric line-of-sight velocities $v_{\textrm{h}}^{\textrm{los}}\pm \Delta \tilde{v}_{\textrm{h}}^{\textrm{los}}$ corresponding to objects towards different Galactic coordinates $(\ell,b)$ and add a random motion of $20\,$km/s in quadrature to $\Delta \tilde{v}_{\textrm{h}}^{\textrm{los}}$ following Sec.~5.1 of Ref.~\cite{DemersBattinelli2007}. The Galactocentric radii $\tilde{R}$ are converted to heliocentric distances $d$ by inverting Eq.~\eqref{eq:R} using $R_0=7.62\,$kpc. We exclude from the full data set star no.~712 (probably belongs to the Sagittarius dwarf galaxy; cf.~Sec.~4).

\subsection{Masers}

\subparagraph{Reid+ '14 \cite{Reid2014}} From Tab.~1, we take the parallaxes $\pi\pm \Delta \pi$, line-of-sight velocities $v_{\textrm{lsr,0}}^{\textrm{los}}\pm \Delta \tilde{v}_{\textrm{lsr}}^{\textrm{los}}$ and equatorial proper motions $\mu_{\alpha^\ast}\pm \Delta \mu_{\alpha^\ast}$, $\mu_\delta\pm \Delta \mu_\delta$ corresponding to objects towards different equatorial coordinates $(\alpha,\delta)$ (then converted to Galactic coordinates $(\ell,b)$) and add a virial motion of $7\,$km/s in quadrature to $\Delta \tilde{v}_{\textrm{lsr}}^{\textrm{los}}$ following Sec.~3 in Ref.~\cite{Reid2009}. The parallaxes are converted to distances through $d/\textrm{kpc}=\textrm{mas}/\pi$, while the equatorial proper motions are converted to Galactic proper motions $\mu_{\ell^\ast}\pm \Delta \mu_{\ell^\ast}$, $\mu_b\pm \Delta \mu_b$. The reported velocities $v_{\textrm{lsr,0}}^{\textrm{los}}$ correspond to a peculiar solar motion $(U,V,W)_{\odot}=(10.3,15.3,7.7)\,$km/s, which is consistently subtracted off in our compilation. We further corrected for the mean peculiar motion of the masers, i.e.~$\bar{U}_s=2.9\,$km/s (cf.~Tab.~4, A5) and $\bar{V}_s=V_\odot-17.1\,$km/s (cf.~Sec.~4.4). We exclude from the full data set eight masers with $R<4\,$kpc (cf.~footnote 3) and 15 outlier masers (cf.~footnote 4).

\subparagraph{Honma+ '12 \cite{Honma2012}} From Tab.~1, we take the parallaxes $\pi\pm \Delta \pi$, line-of-sight velocities $v_{\textrm{lsr,0}}^{\textrm{los}}\pm \Delta \tilde{v}_{\textrm{lsr}}^{\textrm{los}}$ and equatorial proper motions $\mu_{\alpha^\ast}\pm \Delta \mu_{\alpha^\ast}$, $\mu_\delta\pm \Delta \mu_\delta$ corresponding to objects towards different Galactic coordinates $(\ell,b)$ and add a virial motion of $7\,$km/s in quadrature to $\Delta \tilde{v}_{\textrm{lsr}}^{\textrm{los}}$ following Sec.~3 in Ref.~\cite{Reid2009}. The parallaxes are converted to distances through $d/\textrm{kpc}=\textrm{mas}/\pi$, while the equatorial proper motions are converted to Galactic proper motions $\mu_{\ell^\ast}\pm \Delta \mu_{\ell^\ast}$, $\mu_b\pm \Delta \mu_b$. The reported velocities $v_{\textrm{lsr,0}}^{\textrm{los}}$ correspond to a peculiar solar motion $(U,V,W)_{\odot}=(10.3,15.3,7.7)\,$km/s, which is consistently subtracted off in our compilation$^\text{\ref{LSRfootnote}}$. We further corrected for the mean peculiar motion of the masers, i.e.~$\bar{U}_s=2.9\,$km/s (cf.~Tab.~4, A5 in Ref.~\cite{Reid2014}) and $\bar{V}_s=V_\odot-17.1\,$km/s (cf.~Sec.~4.4 in Ref.~\cite{Reid2014}). We exclude from the full data set all objects in (or coincident to objects in) Reid+ '14.

\subparagraph{Stepanishchev \& Bobylev '11 \cite{StepanishchevBobylev2011}} From Tab.~1, we take the parallaxes $\pi\pm \Delta \pi$, heliocentric line-of-sight velocities $v_{\textrm{h}}^{\textrm{los}}\pm \Delta \tilde{v}_{\textrm{h}}^{\textrm{los}}$ and equatorial proper motions $\mu_{\alpha^\ast}\pm \Delta \mu_{\alpha^\ast}$, $\mu_\delta\pm \Delta \mu_\delta$ corresponding to objects towards different equatorial coordinates $(\alpha,\delta)$ (then converted to Galactic coordinates $(\ell,b)$) and add a virial motion of $7\,$km/s in quadrature to $\Delta \tilde{v}_{\textrm{h}}^{\textrm{los}}$ following Sec.~3 in Ref.~\cite{Reid2009}. The parallaxes are converted to distances through $d/\textrm{kpc}=\textrm{mas}/\pi$, while the equatorial proper motions are converted to Galactic proper motions $\mu_{\ell^\ast}\pm \Delta \mu_{\ell^\ast}$, $\mu_b\pm \Delta \mu_b$. We further corrected for the mean peculiar motion of the masers, i.e.~$\bar{U}_s=2.9\,$km/s (cf.~Tab.~4, A5 in Ref.~\cite{Reid2014}) and $\bar{V}_s=V_\odot-17.1\,$km/s (cf.~Sec.~4.4 in Ref.~\cite{Reid2014}). We exclude from the full data set all objects in (or coincident to objects in) Reid+ '14; only one object remains: Ori GMRA.

\subparagraph{Xu+ '13 \cite{Xu2013}} From Tab.~4, we take the parallaxes $\pi\pm \Delta \pi$, line-of-sight velocities $v_{\textrm{lsr,0}}^{\textrm{los}}\pm \Delta \tilde{v}_{\textrm{lsr}}^{\textrm{los}}$ and equatorial proper motions $\mu_{\alpha^\ast}\pm \Delta \mu_{\alpha^\ast}$, $\mu_\delta\pm \Delta \mu_\delta$ corresponding to objects towards different Galactic coordinates $(\ell,b)$ and add a virial motion of $7\,$km/s in quadrature to $\Delta \tilde{v}_{\textrm{lsr}}^{\textrm{los}}$ following Sec.~3 in Ref.~\cite{Reid2009}. The parallaxes are converted to distances through $d/\textrm{kpc}=\textrm{mas}/\pi$, while the equatorial proper motions are converted to Galactic proper motions $\mu_{\ell^\ast}\pm \Delta \mu_{\ell^\ast}$, $\mu_b\pm \Delta \mu_b$. The reported velocities $v_{\textrm{lsr,0}}^{\textrm{los}}$ correspond to a peculiar solar motion $(U,V,W)_{\odot}=(10.3,15.3,7.7)\,$km/s, which is consistently subtracted off in our compilation$^\text{\ref{LSRfootnote}}$. We further corrected for the mean peculiar motion of the masers, i.e.~$\bar{U}_s=2.9\,$km/s (cf.~Tab.~4, A5 in Ref.~\cite{Reid2014}) and $\bar{V}_s=V_\odot-17.1\,$km/s (cf.~Sec.~4.4 in Ref.~\cite{Reid2014}). We exclude from the full data set all objects in (or coincident to objects in) Reid+ '14 and Honma+ '12; only one maser remains: DoAr21/Ophiuchus.

\subparagraph{Bobylev \& Bajkova '13 \cite{BobylevBajkova2013}} From Tab.~1, we take the parallaxes $\pi\pm \Delta \pi$, line-of-sight velocities $v_{\textrm{lsr,0}}^{\textrm{los}}\pm \Delta \tilde{v}_{\textrm{lsr}}^{\textrm{los}}$ and equatorial proper motions $\mu_{\alpha^\ast}\pm \Delta \mu_{\alpha^\ast}$, $\mu_\delta\pm \Delta \mu_\delta$ corresponding to objects towards different equatorial coordinates $(\alpha,\delta)$ (then converted to Galactic coordinates $(\ell,b)$) and add a virial motion of $7\,$km/s in quadrature to $\Delta \tilde{v}_{\textrm{lsr}}^{\textrm{los}}$ following Sec.~3 in Ref.~\cite{Reid2009}. The parallaxes are converted to distances through $d/\textrm{kpc}=\textrm{mas}/\pi$, while the equatorial proper motions are converted to Galactic proper motions $\mu_{\ell^\ast}\pm \Delta \mu_{\ell^\ast}$, $\mu_b\pm \Delta \mu_b$. The reported velocities $v_{\textrm{lsr,0}}^{\textrm{los}}$ correspond to a peculiar solar motion $(U,V,W)_{\odot}=(10.3,15.3,7.7)\,$km/s, which is consistently subtracted off in our compilation$^\text{\ref{LSRfootnote}}$. We further corrected for the mean peculiar motion of the masers, i.e.~$\bar{U}_s=2.9\,$km/s (cf.~Tab.~4, A5 in Ref.~\cite{Reid2014}) and $\bar{V}_s=V_\odot-17.1\,$km/s (cf.~Sec.~4.4 in Ref.~\cite{Reid2014}). We exclude from the full data set all objects in (or coincident to objects in) Reid+ '14, Honma+ '12, Stepanishchev \& Bobylev '11 and Xu+ '13.

\bibliographystyle{apsrev.bst}
\bibliography{galkin}

\end{document}